\documentclass[a4paper,11pt]{article}
\usepackage{amsfonts,amsmath,amssymb}
\usepackage{bm,tensor}
\usepackage{graphicx}
\usepackage{braket}
\usepackage{ifpdf}
\usepackage[utf8]{inputenc}
\usepackage{color}
\usepackage{physics}
\usepackage{url}
\usepackage{jcappub}
\usepackage{mathtools}
\usepackage{subfigure}
\usepackage[colorlinks=true
,urlcolor=blue
,anchorcolor=blue
,citecolor=blue
,filecolor=blue
,linkcolor=blue
,menucolor=blue
,linktocpage=true
,pdfproducer=medialab
,pdfa=true
]{hyperref}

\newcommand*{\DAlambert}{\!\mathop{}\mathbin{\Box}}

\definecolor{dgreen}{rgb}{0.05, 0.70, 0.05}
\definecolor{dblue}{rgb}{0.50, 0.05, 0.50}

\title{Free Streaming Length of Axion-Like Particle After Oscillon/I-ball Decays}
\author[a]{Kaname Imagawa,}
\author[a,b]{Masahiro Kawasaki,}
\author[a]{Kai Murai,}
\author[a]{Hiromasa Nakatsuka,}
\author[a]{and Eisuke Sonomoto}

\affiliation[a]{ICRR, The University of Tokyo, Kashiwa, 277-8582, Japan}
\affiliation[b]{Kavli IPMU(WPI), UTIAS, University of Tokyo, Kashiwa, 277-8583, Japan}

\emailAdd{a0142165@icrr.u-tokyo.ac.jp}
\emailAdd{kawasaki@icrr.u-tokyo.ac.jp}
\emailAdd{kmurai@icrr.u-tokyo.ac.jp}
\emailAdd{hiromasa@icrr.u-tokyo.ac.jp}

\abstract{Axion-like particles (ALPs) are pseudoscalar bosons predicted by string theory.
The ALPs have a shallower potential than a quadratic one, which induces the instability and can form the solitonic object called oscillon/I-ball.
Although the lifetime of oscillons can be very long for some type of potentials, they finally decay until the present.
We perform the numerical lattice simulations to investigate the decay process of oscillons and evaluate the averaged momentum of ALPs emitted from the oscillon decay.
It is found that, if oscillons decay in the early universe, the free-streaming length of ALPs becomes too long to explain the small-scale observations of the matter power spectrum.
We show that oscillons with long lifetimes can change the density fluctuations on small scales, which leads to stringent constraints on the ALP mass and the oscillon lifetime.
}

\keywords{Axion-like particle, Free streaming length, Matter power spectrum, Oscillon, Small-scale crisis}

\arxivnumber{1234.5678}

\begin{document}

\maketitle


\section{Introduction}

The $\Lambda$CDM model, which contains cold dark matter and cosmological constant, explains the cosmological observations very well for scales above $\mathcal{O}$(Mpc)~\cite{Hinshaw_2013,DES:2017myr,Planck:2018vyg}. 
However, below $\mathcal{O}$(Mpc), some observations seem inconsistent with the $\Lambda$CDM model, which is called the small-scale crisis~\cite{Moore:1999nt,de_Blok_2010,Boylan_Kolchin_2012,Bullock:2017xww}.
As a candidate to solve this crisis, fuzzy dark matter\ (FDM), dark matter with ultralight mass, has been attracting much attention~\cite{Hu:2000ke,Hui:2016ltb}.
In fact, FDM with a mass of about $10^{-22}$~eV, whose de Broglie wavelength is $\mathcal{O}$(kpc), smears out the fluctuations on small scales and provides good agreement with the observations on small scales.

Axion-like particle (ALP) is one of the candidates of FDM.
ALP is predicted by string theory and its mass is naturally small because of its shift symmetry~\cite{Svrcek:2006yi}.
The ALP is often considered to have a cosine potential on the analogy of the QCD axion.
However, the ALP can have a different potential in general~\cite{Dong:2010in,Kallosh:2013hoa,Kallosh:2013yoa}.
If the ALP potential is shallower than a quadratic one, the ALP field can be spatially destabilized and non-topological solitons called oscillons or I-balls are formed~\cite{Gleiser:1993pt,Copeland:1995fq,Amin:2011hj,Amin:2010dc,Amin:2019ums,Lozanov:2017hjm,Kitajima:2018zco,Hong:2017ooe}.

Oscillons are quasi-stable and have long lifetimes due to the adiabatic invariance which is approximately conserved.
Oscillons slowly lose their energy emitting scalar radiation.
In this paper, we call this process ``evaporation'' of oscillons.
The evaporation rate is calculated perturbatively~\cite{Ibe:2019vyo,Zhang:2020bec}.
In this quasi-stable phase, oscillons have a configuration in which the field oscillates periodically and the oscillation frequency slowly increases with time as the evaporation proceeds.
When the oscillation frequency reaches a certain value, the oscillon decays at once and the spatially localized energy is released in a very short time compared to its lifetime.
We call this process non-perturbative decay or simply ``decay'' of oscillons, which is mainly investigated in this paper.
When the oscillons emit ALPs with large velocities through the decay, 
their free streaming washes out small-scale density fluctuations,
which could lead to a stringent constraint on the ALPs by observations of the matter power spectrum.
If ALPs account for the dark matter of the universe, a sizable ratio of ALPs forms oscillons, and hence it is important to investigate the evolution of the ALPs taking into account the effects of oscillons.

In this paper, we investigate the effects of the free streaming of ALPs emitted from the oscillon decay on the matter power spectrum.
To this end, we evaluate the momentum distribution of ALPs at the time of oscillon decays and free-streaming length (FSL) using lattice simulations.
Even when ALPs constitute 100\% of the dark matter, not all the ALPs form oscillons and some ALPs remain behaving as cold dark matter~\cite{Kawasaki:2020jnw}.
Therefore, just after the oscillon decays, dark matter is composed of warm (or hot) ALPs emitted from oscillons, cold ALPs, and the other cold dark matter (if any).
This situation is similar to a two-component system containing cold and warm dark matter, which may smooth out the small-scale structures depending on the FSL and abundance of the warm component.
Based on this similarity, we study the matter power spectrum of the universe, including ALPs from oscillons and cold dark matter.
As a result, when the oscillons decay at a temperature below $\mathcal{O}$(1)~eV, we obtain stringent constraints on the ALP mass and the oscillon lifetime from the observations of CMB, large-scale structure of the universe, and Lyman-$\alpha$ forest.

This paper is organized as follows.
In Sec.~\ref{Oscillons of Axion-like Particle}, we derive the initial profile of an oscillon from its basic properties and formulate the ALP emission from the oscillon.
In Sec.~\ref{Simulation}, we perform numerical simulations on the time evolution of the oscillon and estimate the ALP velocity from the oscillon decay.
In Sec.~\ref{Features of Oscillon Decay}, we limit the lifetime and mass of the oscillon from the results obtained in Sec.~\ref{Simulation}.
Sec.~\ref{Conclusion} is devoted to conclusion and discussion.
All cosmological parameters in this paper are extracted from the results of Planck 2018~\cite{Planck:2018vyg}.

\section{Oscillons of Axion-like Particle}
\label{Oscillons of Axion-like Particle}

In this section, we review the properties of oscillons. 
In Sec.~\ref{Oscillon Profile}, we derive the equations that determine the stationary solution of the oscillon profile, which is used for the initial condition in the lattice simulation in Sec.~\ref{Simulation}, based on Refs.~\cite{Kasuya:2002zs,Kawasaki:2015vga}. 
In Sec.~\ref{Evaporation of Oscillons}, we explain the evaporation of oscillons which is described perturbatively.
In Sec.~\ref{Decay of Oscillons}, we introduce the decay of oscillons. In this stage, the spatially localized energy is released in a very short time compared to its lifetime.
We also describe how to calculate the physical quantities such as the momentum distribution of the radiation from the oscillons calculated in the lattice simulation.

\subsection{Oscillon Profile}
\label{Oscillon Profile}

We consider the ALP field $\phi$ whose action in the Minkowski space-time is written as
\begin{equation}
    S 
    =
    \int \mathrm{d}^4 x \,
    \left[ 
        \frac{1}{2} \partial_{\mu} \phi \partial^{\mu} \phi -  V(\phi)
    \right].
\end{equation} 
Here we do not consider the cosmic expansion and the validity of this assumption is discussed in Sec.~\ref{Free-Streaming Length of Axion-like particle}.
We decompose the ALP potential $V(\phi)$ as
\begin{equation}
    V(\phi) = \frac{1}{2} m^2 \phi ^2 + V_{ \mathrm{nl} }(\phi),
    \label{ALPpotential}
\end{equation}
where $m$ is the ALP mass and $V_{\mathrm{nl}}(\phi)$ represents the higher order terms.
In this paper, we adopt the pure natural type potential derived from the pure Yang-Mills gauge theory~\cite{Silverstein:2008sg,McAllister:2008hb,Nomura:2017ehb}
\begin{equation}
    V(\phi) 
    =
    \frac{ m^2M^2 }{2p}
    \left[
        1-\left( 1 + \frac{\phi^2}{M^2} \right)^{-p} 
    \right],
    \hspace{1cm}
    (p>-1),
    \label{ALPpot}
\end{equation}
where $M$ is the effective decay constant of the ALP.
This potential is shallower than a quadratic one for $p>-1$, leading to the formation of oscillons~\cite{Kawasaki:2019czd,Hong:2017ooe}. 

We briefly derive the differential equations that the oscillon profile should satisfy following Refs.~\cite{Kasuya:2002zs,Kawasaki:2015vga}.
Since oscillons show an oscillatory behavior, we write the oscillon configuration as
\begin{equation}
    \phi_{ \mathrm{osc} }( t,\bm{x} )
    =
    2\psi(\bm{x}) \cos{ (\omega_{\mathrm{osc}} t) },
    \label{Oscillon_sol}
\end{equation}
where $\omega_{\mathrm{osc}}$ is the internal frequency of oscillons.
Oscillons have an approximately conserved quantity called an adiabatic invariant which is written as
\begin{equation}
    I 
    =
    \frac{1}{ \omega_{\mathrm{osc}} }
    \int \mathrm{d}^3 x \, \overline{ \dot{\phi} ^2  _{ \mathrm{osc} } },
\end{equation}
where $\overline{ \dot{\phi} ^2 _{\mathrm{osc}} }$ represents the time average of $\dot{\phi} ^2 _{\mathrm{osc}}$ over the oscillation period $2 \pi / \omega_{ \mathrm{osc} }$.
The equation of motion for  $\phi$ is
\begin{equation}
    \frac{\partial^2 \phi}{\partial t^2} 
    - \nabla^2 \phi
    + m^2 \phi
    + \frac{ \partial V_{\mathrm{nl}} }{ \partial \phi }(\phi)
    =0.
    \label{KG}
\end{equation}
Substituting Eq.~\eqref{Oscillon_sol} into Eq.~\eqref{KG}, we obtain
\begin{equation}
    - \omega_{\mathrm{osc}}^2 \psi \cos( \omega_{\mathrm{osc}} t ) 
    - \nabla^2 \psi \cos ( \omega_{\mathrm{osc}} t ) 
    + m^2 \psi \cos ( \omega_{\mathrm{osc}} t ) 
    + \frac{1}{2} \frac{ \partial V_{\mathrm{nl}} }{ \partial \phi } ( 2 \psi \cos \left[ \omega_{\mathrm{osc}} t\right]) 
    = 0.
    \label{KG2}
\end{equation}
Multiplying Eq.~\eqref{KG2} by $\cos (\omega_{\mathrm{osc}}t)$ and taking the average with respect to time, we obtain
\begin{equation}
    \left(
        \frac{ \partial^2 }{ \partial r^2 } + \frac{2}{r} \frac{\partial}{\partial r} 
    \right) \psi 
    =
    ( m^2 - \omega_{ \mathrm{osc} }^2) \psi 
    + \frac{1}{2} \frac{ \partial V_{\mathrm{eff}} }{ \partial \psi } (\psi),
    \label{second}
\end{equation}
where $V_{\mathrm{eff}}(\psi) = \overline{ V_{\mathrm{nl} } (\phi = 2 \psi \cos [\omega_{ \mathrm{osc} } t])}$. 
Here we assume a spherically symmetric configuration $\psi(\bm{x}) = \psi(r)$ and impose the boundary conditions
\begin{equation}
    \lim_{ r\rightarrow 0 } \frac{ \partial \psi(r) }{ \partial r } 
    = \lim_{ r\rightarrow \infty } \psi(r) 
    = 0.
    \label{boundary_condition}
\end{equation}
In order for Eq.~\eqref{second} to have a nonzero solution under these boundary conditions, $\omega_\mathrm{osc}$ should be smaller than a critical value $\omega_c$.
The solution of Eq.~\eqref{second} with the boundary conditions~\eqref{boundary_condition} is used as the initial condition for the lattice simulation.
Considering the evaporation, $\omega_{\mathrm{osc}}$ gradually increases, and the oscillons suddenly decay when $\omega_{\mathrm{osc}}$ reaches $\omega_{ \mathrm{c} }$.

\subsection{Evaporation of Oscillons}
\label{Evaporation of Oscillons}

Oscillons continuously radiate ALPs in the quasi-stable phase~\cite{Mukaida:2016hwd,Eby:2018ufi,Zhang:2020bec}.
We call this phenomenon evaporation of oscillons and also call the emitted radiation perturbative radiation.
The evaporation rate is calculated as follows~\cite{Ibe:2019vyo,Zhang:2020bec}.
The total energy of an oscillon is given by
\begin{eqnarray}
    \label{eq:oscillon_energy}
    E_{\mathrm{tot}} (t) 
    & = &
    \int \mathrm{d} ^3 \bm{x}
    \left[
        \frac{1}{2}\dot{\phi}^2 + (\nabla \phi)^2 + V(\phi)
    \right],
\end{eqnarray}
where the size of the integration region is much larger than the oscillon radius.
Then the evaporation rate is defined as~\cite{Ibe:2019vyo}
\begin{equation}
    \Gamma =\frac{1}{ \bar{E}_{ \rm{tot} } (t) } \left| \frac{ \mathrm{d} }{ \mathrm{d} t } \bar{E}_{ \rm{tot} } (t) \right|.
\end{equation}
The ALP field is decomposed as 
\begin{equation}
    \phi (t, \bm{x}) = \phi_{ \mathrm{osc} } (t, r) + \xi (t, r),
\end{equation}
where $\xi$ is the perturbative radiation of the oscillon.
Here and hereafter, we assume the spherical symmetry.
In the linear order, the equation of motion for $\xi(t, r)$ is given by
\begin{equation}
    \left[
        \DAlambert + V^{''} ( \phi_{ \mathrm{osc} }(t, r) )  
    \right] \xi
    =
    -V^{'}( \phi_{\mathrm{osc}}(r, t) ) 
    + 2 \overline{ V^{'}( \phi_{\mathrm{osc}}(t', r) ) \cos ( \omega_{\mathrm{osc}} t') } 
    \cos ( \omega_{\mathrm{osc}} t),
\end{equation}
where the primes represent the derivative with respect to $\phi$.
We decompose the radiation as $\xi(t, r) = \sum _{j=3} \xi_j (r) \cos ( j \omega_{\mathrm{osc}} t )$ and then the equation of motion for $\xi_j$ is given by
\begin{equation}
    \left[ \nabla ^2 + \kappa_j ^2 \right] \xi_j = - \mathcal{S}_j,
\end{equation}
where $\kappa_j = \sqrt{j^2 \omega^2 - m^2}$ and $\mathcal{S}_j (r)$ is the source term including the combination of $\xi_k$'s~\cite{Zhang:2020bec}.
The radiative solution of $\xi$ for $r \to \infty$ is given as
\begin{eqnarray}
    \xi_j (t,r)
    &\simeq &
    \frac{1}{4\pi r} \sum_{j=3} \tilde{\mathcal{S}}_j (\kappa_j) \cos ( \kappa_j r - j \omega_{\mathrm{osc}} t ),
    \\
    \tilde{ \mathcal{S}_j } (k)
    &=&
    \int \mathrm{d}^3 \bm{x}^{\prime} \mathcal{S}_j (r^{\prime}) e^{ -i \bm{k} \cdot \bm{x}^{\prime} }.
\end{eqnarray}
Thus the time-averaged classical energy loss rate is obtained by 
\begin{equation}
    \frac{ \mathrm{d} \bar{E}_{ \mathrm{tot} } (t)}{ \mathrm{d} t }
    =
    4 \pi r^2 \overline{ \partial_{t} \xi (t, r) \partial_{r} \xi (t, r) }
    =
    -\frac{1}{8\pi} \sum_{j=3} j \omega_{ \mathrm{osc} } \kappa_j \tilde{\mathcal{S}}_j ( \kappa_j )^2.
\end{equation}

\subsection{Decay of Oscillons}
\label{Decay of Oscillons}

In this subsection, we explain the decay of oscillons and introduce some physical quantities to investigate particle emission from oscillons.
Oscillons emit a large amount of radiation at the end of their life, releasing all of their energy promptly.
We call this phenomenon oscillon decay and also call the particle emission non-perturbative  particle emission or simply particle emission.

We first describe the ALP emission to find the energy distribution for each momentum mode
and the Poynting vector that represents the ALP energy flow.
The classical ALP field in momentum space is written as
\begin{equation}
    \xi(t,\bm{x}) 
    =
    \int \frac{\mathrm{d}^3 \bm{k}}{\sqrt{2\omega_k (2\pi)^3}} \, 
    \left[ 
        \xi_{\bm{k}} e^{ i\omega_k t - i \bm{k} \cdot \bm{x} } 
        + \xi^{*}_{\bm{k}} e^{ -i\omega_k t + i \bm{k} \cdot \bm{x} }
    \right],
    \label{Expansion3}
\end{equation}
where $\xi_{\bm{k}}$ is a complex coefficient and  $\omega_{k} = \sqrt{ |\bm{k}|^2 + m^2 }$.
On the other hand, in the lattice simulation, we obtain the Fourier transform of the field value $\xi(t, \bm{x}_0)$ with respect to the time as 
\begin{equation}
    \xi( t, \bm{x}_0 )
    =
    \int_{-\infty}^{\infty} \mathrm{d}\omega \, \xi_\omega (\bm{x}_0) e^{ i\omega t},
    \label{timeFourier3}
\end{equation}
where $\bm{x}_0$ is some fixed point and we take this point to be far enough away from an oscillon.
To compare the two formulas, we rewrite the equation as follows
\begin{eqnarray}
    \xi (t, \bm{x}_0 ) 
    &=&
    \int_{0}^{\infty} \mathrm{d} \omega
    \left[ 
        \xi_{\omega} (\bm{x}_0) e^{ i \omega t}
        + \xi^{*}_{\omega} (\bm{x}_0) e^{ -i \omega t} 
    \right]
    \label{timeFourier4}
    \\
    & \simeq &
    \int_{0}^{\infty} \left( \mathrm{d} k \frac{k}{ \omega _k } \right)
    \left[ 
        \xi_{\omega} (\bm{x}_0) e^{ i \omega _{k} t}
        + \xi^{*}_{\omega} (\bm{x}_0) e^{ -i \omega _{k} t} 
    \right]
    \label{timeFourier5}
    \\
    &=&
    \int \frac{ d^3 \bm{k} }{ \sqrt{ 2\omega_{k} ( 2\pi )^3} }
    \left[ 
        \sqrt{ \frac{\pi}{ k^2 \omega_k } } \xi_{\omega} (\bm{x}_0) e^{ i \omega _{k} t}
        +  \sqrt{ \frac{\pi}{ k^2 \omega_k } } \xi^{*}_{\omega} (\bm{x}_0) e^{ -i \omega _{k} t} 
    \right],
    \label{timeFourier6}
\end{eqnarray}
where, in Eq.~\eqref{timeFourier4}, we use $\xi^{*} _{\omega} = \xi_{-\omega}$, and, in Eq.~\eqref{timeFourier5}, we use the on-shell condition $\omega_{k} = \sqrt{ |\bm{k}|^2 + m^2 }$ to rewrite the measure and neglect the range of integration from $0$ to $m$.
Comparing Eq.~\eqref{timeFourier6} with Eq.~\eqref{Expansion3}, we get
\begin{equation}
    |\xi_{k}| = \sqrt{ \frac{\pi}{ k^2 \omega_k } } | \xi_{\omega_k} |,
    \label{relation_of_xi}
\end{equation}
which connects the analytical field values and numerical values obtained by lattice simulations.
We use the spherical symmetry to drop the $\bm{k}$-direction dependence of $\xi_{\bm{k}}$.
Here and hereafter, we omit the $\bm{x}$-dependence of $\xi_{\omega}$.

Let us consider the non-perturbatively emitted energy from oscillons for each momentum mode after the decay of oscillons.
In this stage, the solitonic solution~\eqref{Oscillon_sol} no longer holds, and the energy is promptly released (see also Sec.~\ref{Simulation Results}).
Therefore, after the oscillon decay, we can ignore the field value of oscillons and only consider radiation.
The total energy of the radiation is
\begin{eqnarray}
    E_{\mathrm{rad}} 
    &= &
    \int \mathrm{d}^3 \bm{x} \, 
    \left[ 
        \frac{1}{2}\dot{\xi}^2+\frac{1}{2}(\nabla \xi)^2+\frac{1}{2}m^2\xi^2
    \right] 
    \label{energy2} 
    \\
    &=&
    \int \mathrm{d}^3 \bm{k} \, \omega_k |\xi_k|^2,
    \label{energy1} 
\end{eqnarray}
where $k=|\bm{k}|$.
Here, we assume $|\phi/M|$ is small enough to neglect the non-linear terms of $V(\phi)$ in estimating $E_\text{rad}$.
We numerically confirmed that the condition $|\phi/M| \ll 1$ is satisfied even immediately after the oscillon decay.
From Eqs.~\eqref{energy1} and \eqref{relation_of_xi}, we have
\begin{equation}
    \frac{ \mathrm{d}E_{\mathrm{rad}} }{\mathrm{d}\ln{k}} 
    =
    4\pi k^3 \omega_k |\xi_k|^2
    =
    4 \pi ^2 k |\xi_{\omega}|^2.
    \label{energy_distribution0}
\end{equation}
This formula~\eqref{energy_distribution0} is valid for both perturbative radiation and non-perturbative particle emission.

Now we can calculate the average momentum when the ALPs are emitted from oscillons.
The average momentum is given by
\begin{equation}
    \bar{k} 
    =
    \frac{\int \mathrm{d}^3 k \, n_k k}{\int \mathrm{d}^3 k \, n_k}
    =
    \frac{ \int_0^{\infty} \mathrm{d} k \, 4 \pi^2 k |\xi_{\omega}|^2 / \omega_k }{ \int_0^{\infty} \mathrm{d} k \, 4 \pi^2 |\xi_\omega|^2 / \omega_k},
    \label{averaged_momentum}
\end{equation}
where $n_k = E_k / \omega_k$ and $E_k$ is the Fourier mode of the radiation energy as $E_{ \mathrm{rad} } = \int \mathrm{d}^3 \bm{k} \, E_k$. 

We evaluate the Poynting vector  near the boundary to see the evolution of the energy of the outgoing ALPs.
Since, at the position far from the origin, the oscillon profile $\phi_{ \mathrm{osc} }$ is negligibly small, the Poynting vector is written as 
\begin{equation}
    \frac{1}{4 \pi r^2} \frac{ \mathrm{d} E_{ \mathrm{tot} } (t) }{ \mathrm{d} t }
    =
    T_{0r} (t) 
    \simeq 
    \frac{ \partial \xi(r, t) }{\partial r}
    \frac{ \partial \xi(r, t) }{\partial t}  .
    \label{numerical_Poynting}
\end{equation}

\section{Simulation}
\label{Simulation}

We study the free streaming of the ALPs released from the non-perturbative decay of oscillons and its effect on the density fluctuations.
For this purpose, we evaluate the momenta of the ALPs emitted at the time of the non-perturbative decay of oscillons using lattice simulations.

\subsection{Simulation Setup}

Imposing the spherical symmetry, the equation of motion for the ALP field $\phi$ is given by
\begin{equation}
    \left(
        \frac{\partial^2}{\partial t^2} - \frac{\partial^2}{\partial r^2}
        - \frac{2}{r}\frac{\partial}{\partial r} + m^2
    \right) \phi 
    = 
    -\frac{ \partial V_{\mathrm{nl}}(\phi) }{ \partial \phi },
\end{equation}
which we solve numerically.
In the lattice simulation, we normalize the field and coordinates by $m$ and $m^{-1}$ as
\begin{equation}
    \phi\rightarrow m\phi,\quad  t\rightarrow \frac{t}{m},\quad  x\rightarrow\frac{x}{m}.
\end{equation}
The simulation parameters are summarized in Table~\ref{paramater}.
The box size is chosen to be sufficiently large compared to the radius of the oscillon ($\sim 10m^{-1}$).

\begin{table}[t]
    \centering
    \begin{tabular}{cc}
        \hline\hline
        $\omega_{\mathrm{osc}}|_{t=0}$&varying\\
        Box size $L$&64~$m^{-1}$\\
        Grid size $N$&1024\\
        Initial time&0\\
        Final time&$5.0\times 10^5 ~m^{-1}$\\
        Time step&$1.0\times 10^{-2}~m^{-1}$\\
        \hline\hline
    \end{tabular}
    \caption{Simulation parameter.}
    \label{paramater}
\end{table}

The initial field values are set by solving Eqs.~\eqref{second} and \eqref{boundary_condition}.
At the initial time $t=0$, we set 
\begin{equation}
    \dot{\phi}(t=0,r) = 0,
\end{equation}
and we always impose
\begin{equation}
    \left.\frac{1}{r}\frac{ \partial \phi (t, r) }{\partial r}\right|_{r=0} = 0.
\end{equation}
At the boundary of the simulation box $r=L$, we impose the absorbing boundary condition, which leads to the absorption of the ALPs emitted from the oscillon~\cite{Salmi:2012ta}.
The time evolution and spatial derivative are described by the fourth-order symplectic integration scheme and fourth-order central difference scheme, respectively.
We confirmed that the simulation results do not significantly depend on the box size $L$, grid size $N$, and time step.

\subsection{Simulation Results}
\label{Simulation Results}

We numerically calculate the total energy in the simulation box and Poynting vector using Eqs.~\eqref{eq:oscillon_energy} and \eqref{numerical_Poynting}, respectively.
\begin{figure}
    \subfigure{%
        \includegraphics[width=7.0cm]{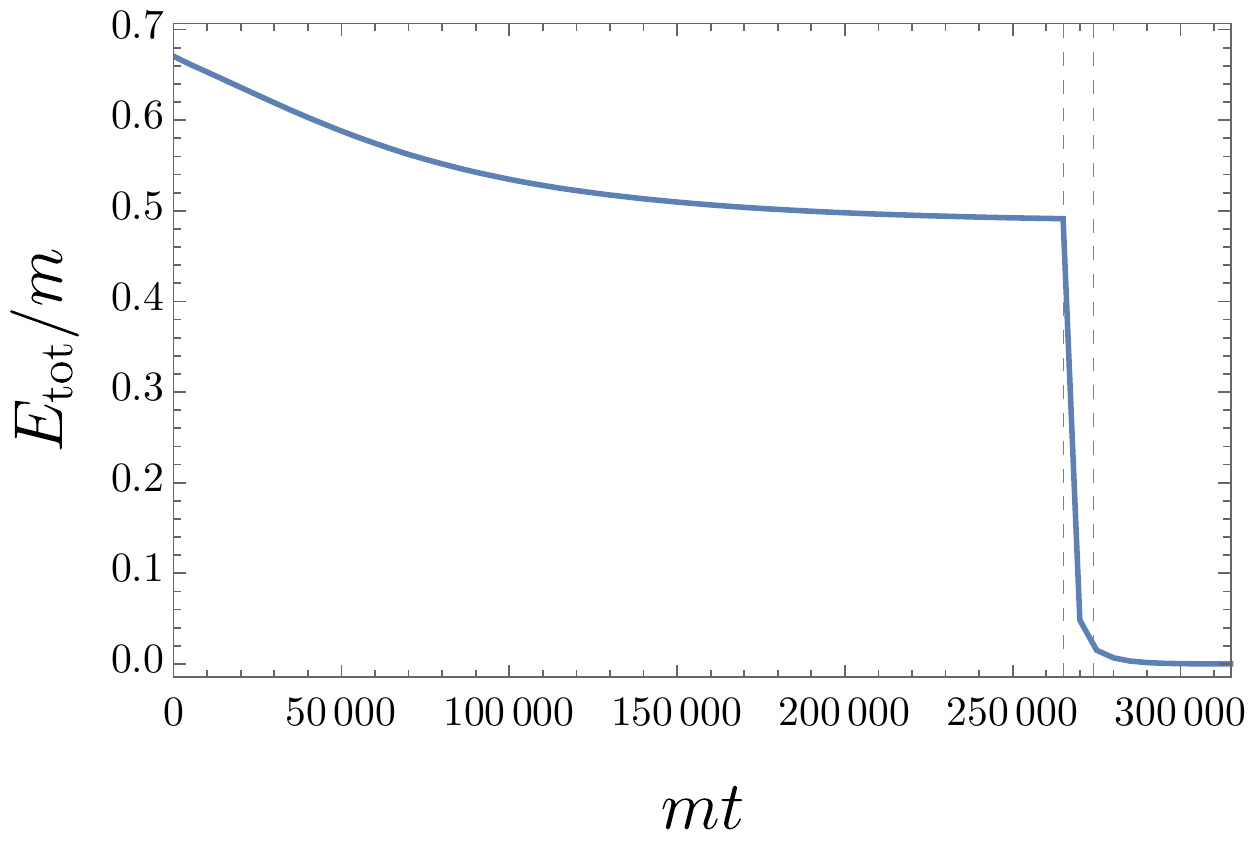}}%
    \subfigure{%
        \includegraphics[width=7.3cm]{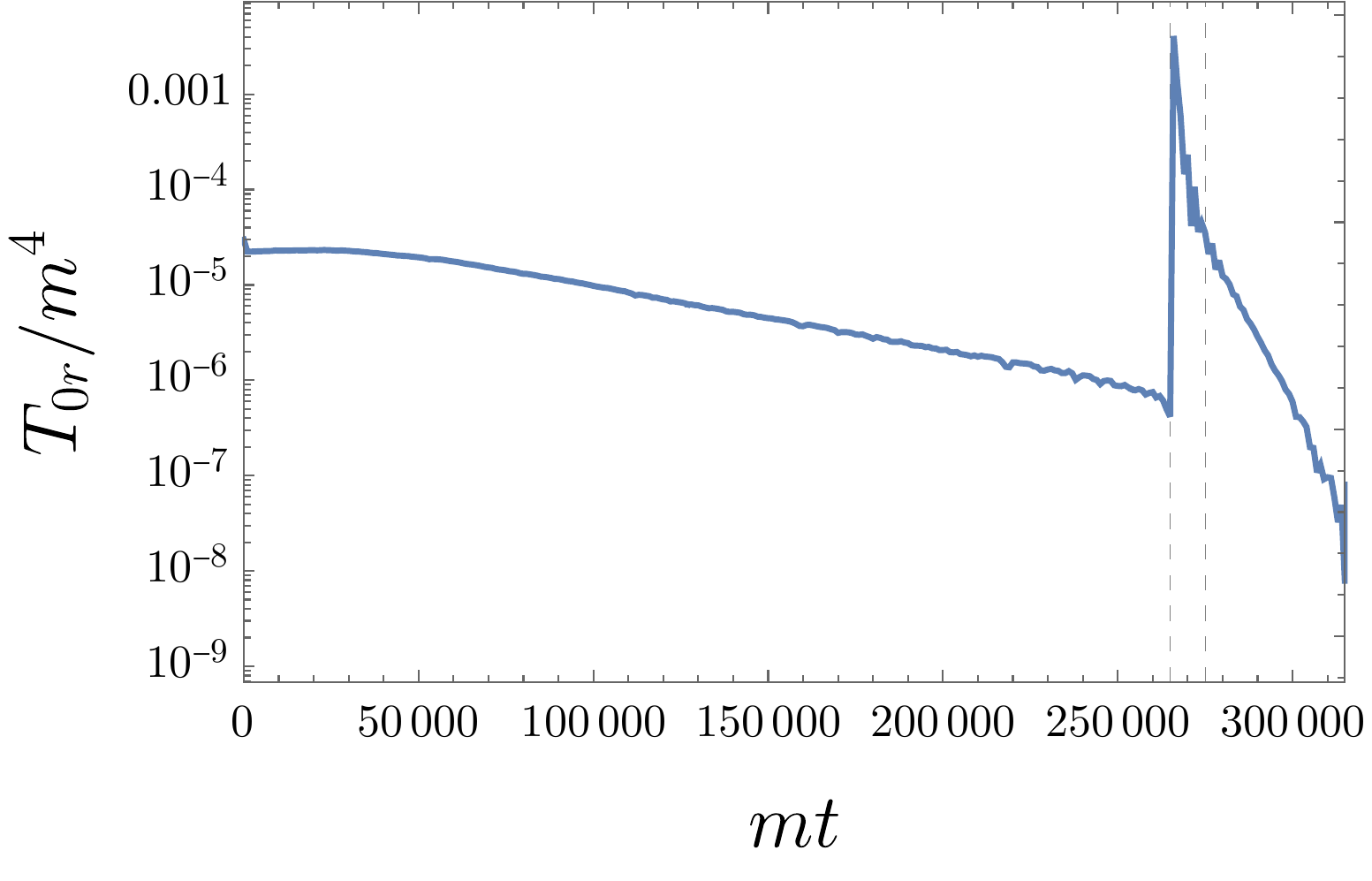}}%
    \caption{
        Time evolution of the energy and Poynting vector. 
        The left panel shows the time evolution of the total energy in the simulation box.
        Note that, since we are using the absorbing boundary condition, the total energy is not conserved.
        The right panel shows the time evolution of the Poynting vector at $r=60m^{-1}$.
        The Poynting vector is averaged over a time interval $10^3 m^{-1}$.
        The gray vertical dashed lines separate the three regions: before, immediately after, and after the oscillon decays.
        The gray lines of the left and right represent the times when the oscillon decays and when 95\% of the energy is lost respectively. 
        Eq.~\eqref{ALPpot} is used for the potential with $p=2$.
        }
    \label{Energy_and_Poynting} 
\end{figure}
The left panel of Fig.~\ref{Energy_and_Poynting} shows the time evolution of the energy inside the simulation box,
and the right panel of Fig.~\ref{Energy_and_Poynting} shows the time-averaged Poynting vector near the boundary.
From the left panel of Fig.~\ref{Energy_and_Poynting}, we can see that there are three stages for the time evolution of the system.
During the early stage, $t \lesssim 2.6 \times 10^5 m^{-1}$, the oscillon is almost stable and gradually releases its energy as radiation.
In the transient stage $2.6 \times 10^5 m^{-1} \lesssim t \lesssim 2.7 \times 10^5 m^{-1}$, the oscillon decays non-perturbatively and releases most of its energy.
After the oscillon decays $t \gtrsim  2.7 \times 10^5 m^{-1}$, the value of the Poynting vector in the right panel of Fig.~\ref{Energy_and_Poynting} decreases rapidly.
From Fig.~\ref{Energy_and_Poynting}, we can see 
\begin{equation}
    \frac{t_\text{end}-t_{ \rm{decay} }}{ t_{ \rm{decay} }}
    =\frac{ \Delta t_{ \rm{decay} }}{ t_{ \rm{decay} }}
    \simeq
    \frac{ 10^4 m^{-1} }{ 3 \times 10^5 m^{-1} }
    \simeq
    0.03,
    \label{decay_time_interval}
\end{equation}
where $t_{ \mathrm{decay} }$ is the decay time, $t_{ \mathrm{end} }$ is the time when 95 \% of the energy is lost and $\Delta t_\text{decay}=t_\text{end}-t_\text{decay}$.
Thus, an oscillon releases most of its energy in a very short time scale compared to its lifetime.

\begin{figure}[bhtp]
    \begin{center}
        \includegraphics[clip,width=12.5cm]{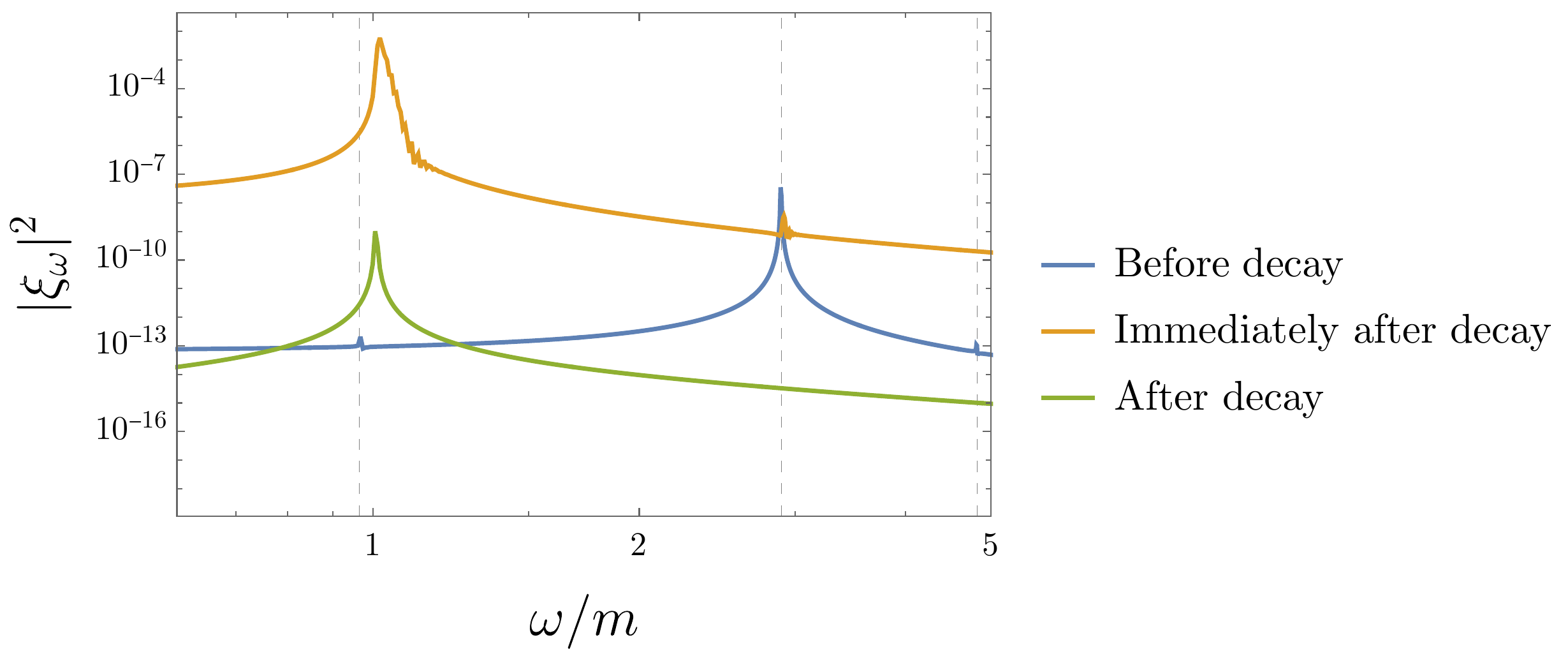}
        \caption{
        Fourier transform of $\xi(t,r=60m^{-1})$.
        The vertical axis is proportional to the energy of the particle with each frequency (see Eq.~\eqref{energy_distribution0}).
        Blue, green, and orange lines correspond to
        the results calculated in different time intervals of $[2.50,\, 2.51]$, $[2.66,\, 2.67]$, and $[3.50,\, 3.51]$ (in units of $10^5 m^{-1}$), respectively.
        Gray dashed lines represent $\omega \simeq \omega_{\mathrm{osc}}, 3\omega_{\mathrm{osc}}$, and $5\omega_{\mathrm{osc}}$, respectively with $\omega_\mathrm{osc}$ at $t = 2.50 \times 10^5 m^{-1}$.
        }
        \label{Fourier_transform}
    \end{center}
\end{figure}

In order to obtain the energy distribution of the emitted ALPs immediately after the oscillon decay, we perform the Fourier transform of $\xi(t,r\sim L)$ with respect to time and obtain $\xi_{\omega}$ [see Eq.~\eqref{timeFourier3}].
Since we use the field values near the boundary of the simulation box, we can safely neglect the effect of the oscillon profile.
The result is shown in Fig.~\ref{Fourier_transform}. 
The energy distribution of the ALP radiation before the oscillon decay is shown by the blue line in the figure, where the mode with $\omega\sim 3\omega_{\mathrm{osc}}$ is dominant.
This is consistent with the analytic calculation of evaporation~\cite{Zhang:2020bec,Ibe:2019vyo}.
On the other hand, immediately after the oscillon decays, the mode with $\omega\sim 1.1 m$ is dominant as shown by the orange line.
After that, all modes reduce the energy as shown by the green line.

\begin{figure}[t]
    \begin{center}
        \includegraphics[clip,width=11cm]{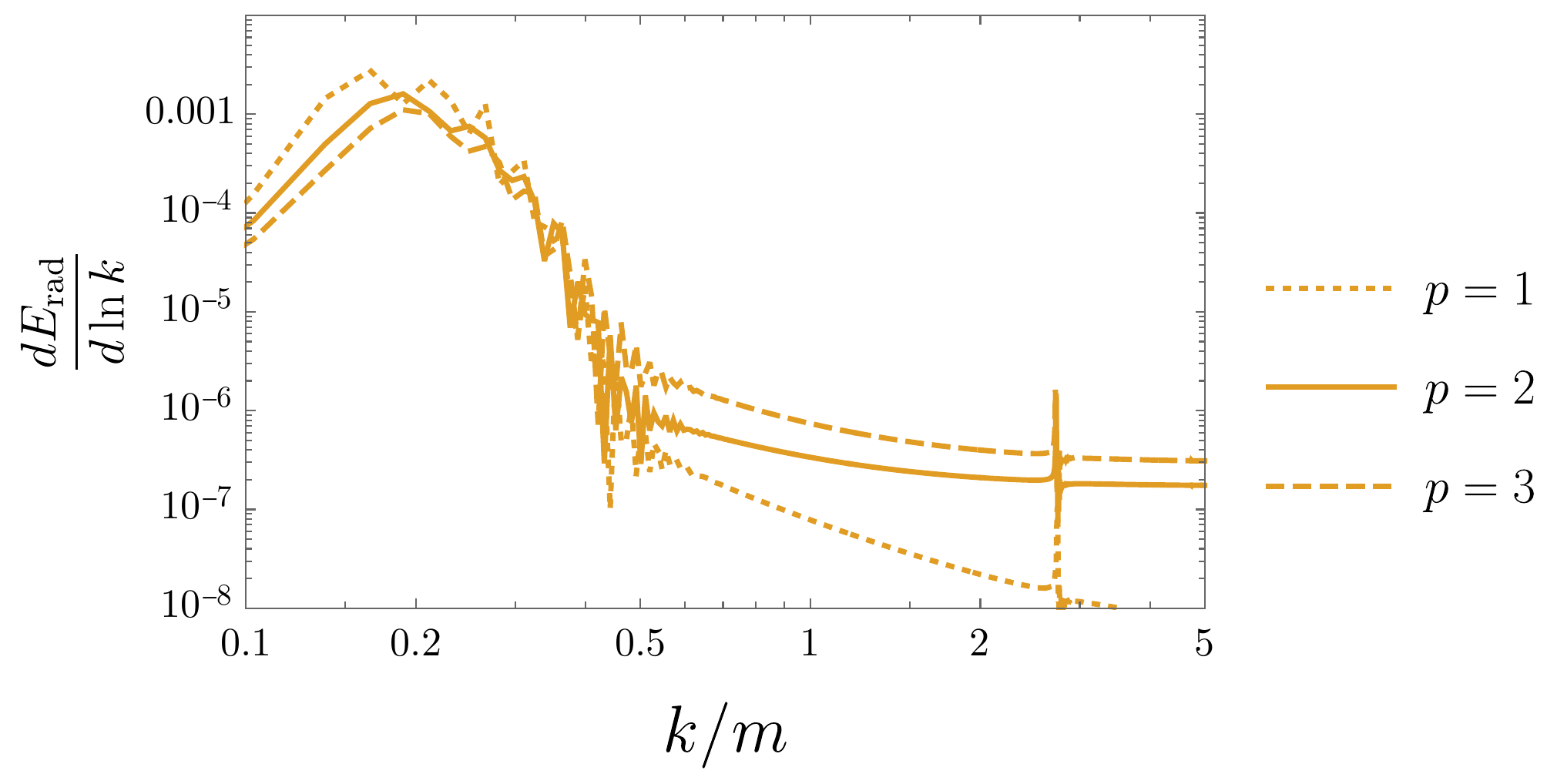}
        \caption{
        Fourier transform of $\xi(t,r=60m^{-1})$.
        The dotted, solid and dashed lines represent the Fourier transform with $p=1$, $2$, and $3$ respectively.
        This time interval with $p=2$ corresponds to that of the orange line in Fig.~\ref{Fourier_transform}.
        We use the on-shell condition $\omega = \sqrt{k^2 + m^2}$ to rewrite the orange line of Fig.~\ref{Fourier_transform} to Fig.~\ref{Fourier_transform2} and also use Eq.~\eqref{energy_distribution0}.
        The spikes at $k/m \sim 3$ are the perturbative radiations from the oscillon with $\omega \simeq 3 \omega_{ \mathrm{osc} }$
        , which are emitted when oscillons are in the quasi-stable phase.
        }
        \label{Fourier_transform2}
    \end{center}
\end{figure}

Next, we evaluate the momentum distribution of the ALP emission using the on-shell condition $\omega = \sqrt{k^2 + m^2}$.
We can safely neglect the mode with $\omega \lesssim m$ since the value of $|\xi_{ \omega }|^2$ with $\omega \lesssim m$ is much smaller than the peak at $\omega \sim 1.1 m$.
Then we rewrite the horizontal axis of Fig.~\ref{Fourier_transform} in terms of the momentum $k$
as shown in Fig.~\ref{Fourier_transform2}.
We can say from Fig.~\ref{Fourier_transform2} that the dominant contribution of the emitted energy is not one with $k/m \sim 3$, which is the radiation emitted in the quasi-stable stage of oscillon, but one with $k/m \sim \mathcal{O}(0.1)$.
From the result in Fig.~\ref{Fourier_transform2} and Eq.~\eqref{averaged_momentum}, we obtain
\begin{eqnarray}
    \bar{k} &\simeq& 0.195m, 
    \label{momentum}
    \\
    \bar{v} &=&\bar{k} / \sqrt{ \bar{k}^2 + m^2 } \simeq 0.191,
    \label{velocity}
\end{eqnarray}
where $\bar{v}$ is the averaged velocity.
From Eq.~\eqref{velocity}, we can see that the velocity of the ALPs emitted in the non-perturbative decay is $\mathcal{O}(10)\%$ of the speed of light.
As we will see later in Sec.~\ref{Matter Power Spectrum of Axion-like particles}, the release of the ALPs with non-negligible momenta can suppress the small-scale density perturbations.

\section{Cosmological Effect of Oscillon Decay}
\label{Features of Oscillon Decay}

\subsection{Free-Streaming Length of Axion-like particles}
\label{Free-Streaming Length of Axion-like particle}

In this subsection, we estimate the free-streaming length (FSL) of the ALPs emitted from the non-perturbative decay of oscillons.
We focus on the time when oscillons release large amounts of their energy in a short period of time.
As shown in the right panel of Fig.~\ref{Energy_and_Poynting},  the Poynting vector during the quasi-stable phase of the oscillon is much smaller than that at the time of the non-perturbative decay.
Therefore, we ignore the free-streaming due to the evaporation.

We assume that the ALPs are instantaneously emitted at $t_\mathrm{decay}$ and that the cosmic expansion is negligible in calculating the value of $\bar{k}$, which is justified if
\begin{eqnarray}
    &&
    \Delta t_{ \mathrm{decay} } 
    \ll
    H( t_{ \mathrm{decay} } )^{-1}
    \label{FSvsExpansion1}
     ~~\Leftrightarrow ~~
    \Delta t_{ \rm{decay} } / t_{ \rm{decay} }
    \ll
    1,
\end{eqnarray}
where $H(t) ^{-1}$ is the Hubble time, and $H( t_{ \mathrm{decay} } ) ^{-1} = 2 t_{ \mathrm{decay} } $ in the radiation dominated era.
Since, from our numerical simulation, we obtained $\Delta t_\text{decay}/t_\text{decay}\simeq 0.03$ as shown in Eq.~\eqref{decay_time_interval}, the condition~\eqref{FSvsExpansion1} is satisfied.

The comoving FSL is calculated as
\begin{align}
    \ell_{\mathrm{FSL}}( t_{ \mathrm{decay} } ) 
    &=
    \int_{ t_{\mathrm{decay}} }^{t_0} \mathrm{d}t \, \frac{v(t)}{a(t)}
    \nonumber \\
    &=
    H_{0}^{-1} \int_{a( t_{ \mathrm{decay} } )}^{a(t_0)} \, 
    \frac{\mathrm{d}\tilde{a}}{\tilde{a}^2 \sqrt{\Omega (\tilde{a})}} 
    \frac{k(\tilde{a})}{\sqrt{k(\tilde{a})^2 + m^2}},
    \label{free-streaming length}
\end{align}
where $a=a(t)$ is the scale factor, $H_{0}$ is the present Hubble parameter, $\Omega(a) = ( \Omega_{\mathrm{r,0}}/a^4 + \Omega_{\mathrm{m,0}}/a^3 + \Omega_{\rm{\Lambda}} ) ^{1/2}$, $\Omega_{i,0}$ (i=r,m,$\Lambda$) is the present density parameters of radiation, matter, and cosmological constant, $v(t)$ is the ALP velocity, and $k(a) = \bar{k} \, a(t_{\mathrm{decay}})/ a$ is the average physical momentum with $\bar{k}$ given by~Eq.~\eqref{momentum}.

\subsection{Matter Power Spectrum of Axion-like particles}
\label{Matter Power Spectrum of Axion-like particles}

We consider the matter power spectrum in the scenario where oscillons decay at the cosmic temperature $T=T_{\mathrm{decay}}$.
For simplicity, we assume that the ALPs account for all dark matter of the universe.
When oscillons are formed, 50\% to 70\% of ALPs are confined in the oscillons~\cite{Kawasaki:2020jnw} and the rest of the ALPs do not form oscillons but behave as cold dark matter.
Therefore, at the time of the oscillon decay, dark matter consists of two components,
warm ALPs emitted from oscillons and cold ALPs that have not formed oscillons.
We also assume that all oscillons decay at the same time as an approximation.
Actually, the produced oscillons have some mass distribution.
However, since the evaporation rate is larger for larger oscillons, oscillons quickly evolve to ones with a smaller size and spend most of their lifetimes. 
Thus, the decays of oscillons occur in the similar redshift.

We adopt the following formula used in the context of warm dark matter to calculate the matter power spectrum~\cite{Kamada:2016vsc,Inoue:2014jka}
\begin{equation}
    \frac{P_{ \mathrm{ALP} }(k) }{ P_{ \Lambda \mathrm{CDM}}(k) }
    =
    ( 1 - f_{ \mathrm{osc} } ) + \frac{f_{ \mathrm{osc} }}{ (1 + k / k'_{ \mathrm{d} })^{0.7441} },
    \label{MPS_foemula}
\end{equation}
where $f_\text{osc}$ and $k_\text{d}'$ are given by
\begin{eqnarray}
    f_{ \mathrm{osc} }(r_{ \mathrm{osc} }) 
    &=&
    1-\exp( -a\frac{ r^b_{ \mathrm{osc} } }{1 - r^c_{ \mathrm{osc} } } ), 
     \label{fittingfunc1}
    \\
    k'_{\rm{d}}(k_{ \mathrm{d} }, r_{ \mathrm{osc} }) 
    &=&
    k_{ \mathrm{d} } / r_{ \mathrm{osc} }^{1/2},
    \label{damping_scale}
    \\
    k_{ \mathrm{d} }( k_{ \mathrm{fs} } ) 
    &=&
    2.206h\, \mathrm{Mpc}^{-1} \left( \frac{ k_{\mathrm{fs} } }{ h \mathrm{Mpc} } \right) ^{1.703},
    \label{fittingfunc2}
\end{eqnarray}
with
\begin{equation}
    (a, b, c) = (1.551, 0.5761, 1.263).
    \label{N body parameter}
\end{equation}
Here $k_{ \mathrm{fs} }( t_{ \mathrm{decay} } ) = 2\pi / \ell_{ \mathrm{FSL} }( t_{ \mathrm{decay} } )$ and $r_{ \mathrm{osc} } = \Omega_{ \mathrm{osc} } / \Omega_{ \mathrm{DM} }$ is the density ratio of the oscillons to all dark matter at the time of the oscillon decay.
$P_{ \Lambda \mathrm{CDM} } (k)$ in Eq.~\eqref{MPS_foemula} is calculated by CAMB~\cite{Lewis:1999bs}.
The formula of the matter power spectrum~\eqref{MPS_foemula} is obtained by N-body simulations for the two-component system of warm and cold dark matter and the parameters~\eqref{N body parameter} are determined by fitting the N-body result to Eq.~\eqref{fittingfunc1}.
Note that the $r_{ \mathrm{osc} }$-dependence of the damping scale $k'_{ \mathrm{d} }$ in Eq.~\eqref{damping_scale} differs from that in Ref.~\cite{Kamada:2016vsc}, where the authors assume the thermal warm dark matter.
In Ref.~\cite{Kamada:2016vsc}, it is presumed that the damping scale $k'_{ \mathrm{d} }$ has the same dependence on $r_{ \mathrm{osc} }$ as the Jeans scale which is written as
\begin{equation}
    k_{ \mathrm{J} } 
    =
    \left.
        a\sqrt{ \frac{4 \pi G \rho_{ \mathrm{M} } }{ \sigma ^2 } }
    \right|_{t=t_{ \mathrm{eq} }},
\end{equation}
where $G$ is the gravitational constant, $\rho_{ \mathrm{M} }$ is the matter energy density, and $\sigma ^2$ is the mass-weighted mean squared velocity of the whole dark matter.
In the present case, $\sigma ^2 = r_{ \mathrm{osc} } \sigma_{ \mathrm{osc} }^2$ with $\sigma_{ \mathrm{osc} } \sim \bar{k} / \sqrt{\bar{k}^2 + m^2} = \bar{v}$.
Since $\bar{v}$ is independent of $t_\text{osc}$,
we have $k_{ \mathrm{J} } \propto 1 / \sqrt{\sigma^2} \propto 1/r_{ \mathrm{osc} }^{1/2}$. 

\begin{figure}[t]
    \begin{center}
        \includegraphics[clip,width=14cm]{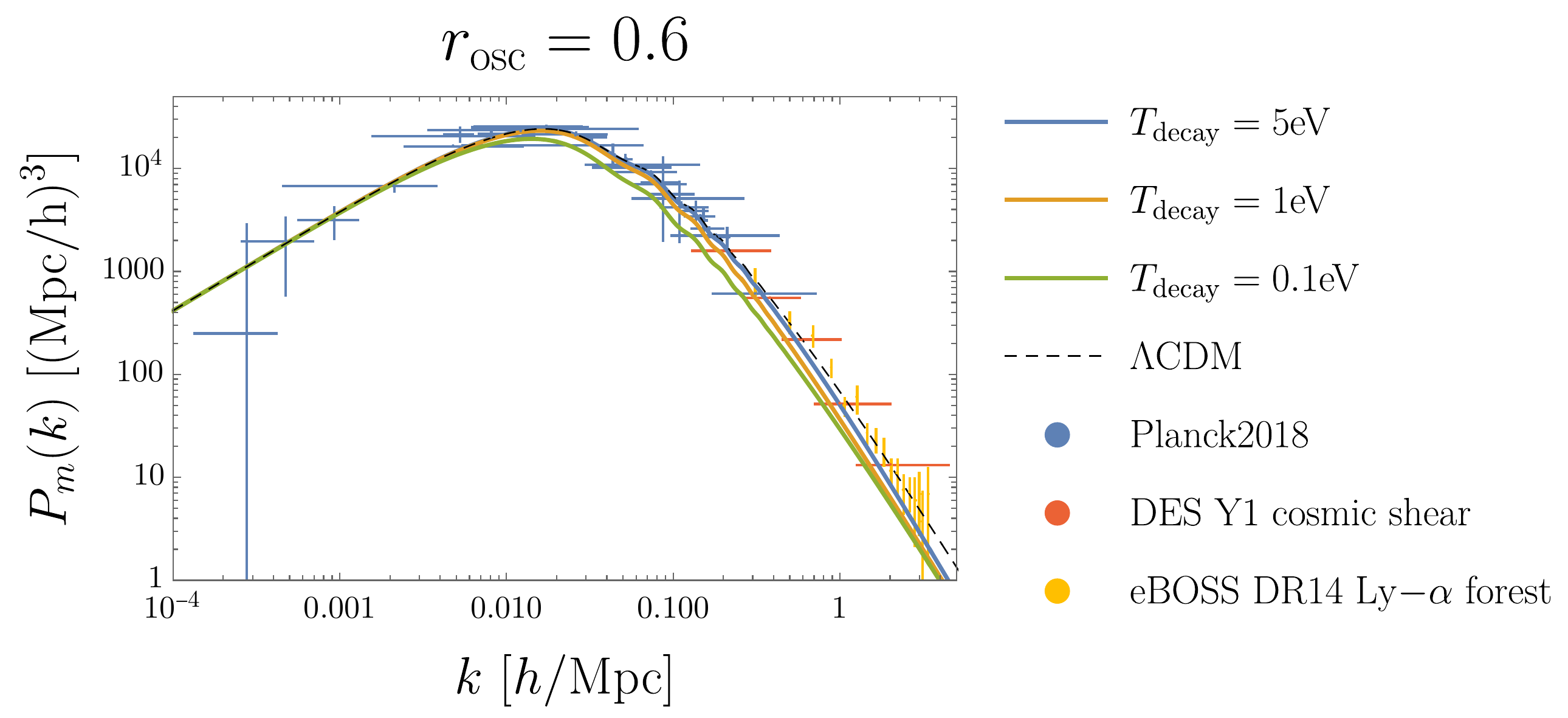}
        \caption{
        Matter power spectrum with $r_{ \mathrm{osc} } = 0.6$.
        The blue, orange and green solid lines represent the matter power spectra with the oscillon decays at the temperature $T_{ \mathrm{decay} } =$ 5 eV, 1 eV, and 0.1 eV, respectively. 
        The black dashed line represents the matter power spectrum in the $\Lambda$CDM model, which is calculated by CAMB~\cite{Lewis:1999bs}.
        The blue, red, and yellow data points with error bars are the observational data of Planck 2018~\cite{Planck:2018vyg}, DES Y1 cosmic shear~\cite{DES:2017qwj}, and eBOSS DR14 Lyman-$\alpha$ forest~\cite{Chabanier:2019eai,2018ApJS..235...42A}, respectively.
        }
        \label{matter_power_spectrum1}
    \end{center}
\end{figure}

\begin{figure}[t]
    \begin{center}
        \includegraphics[clip,width=14cm]{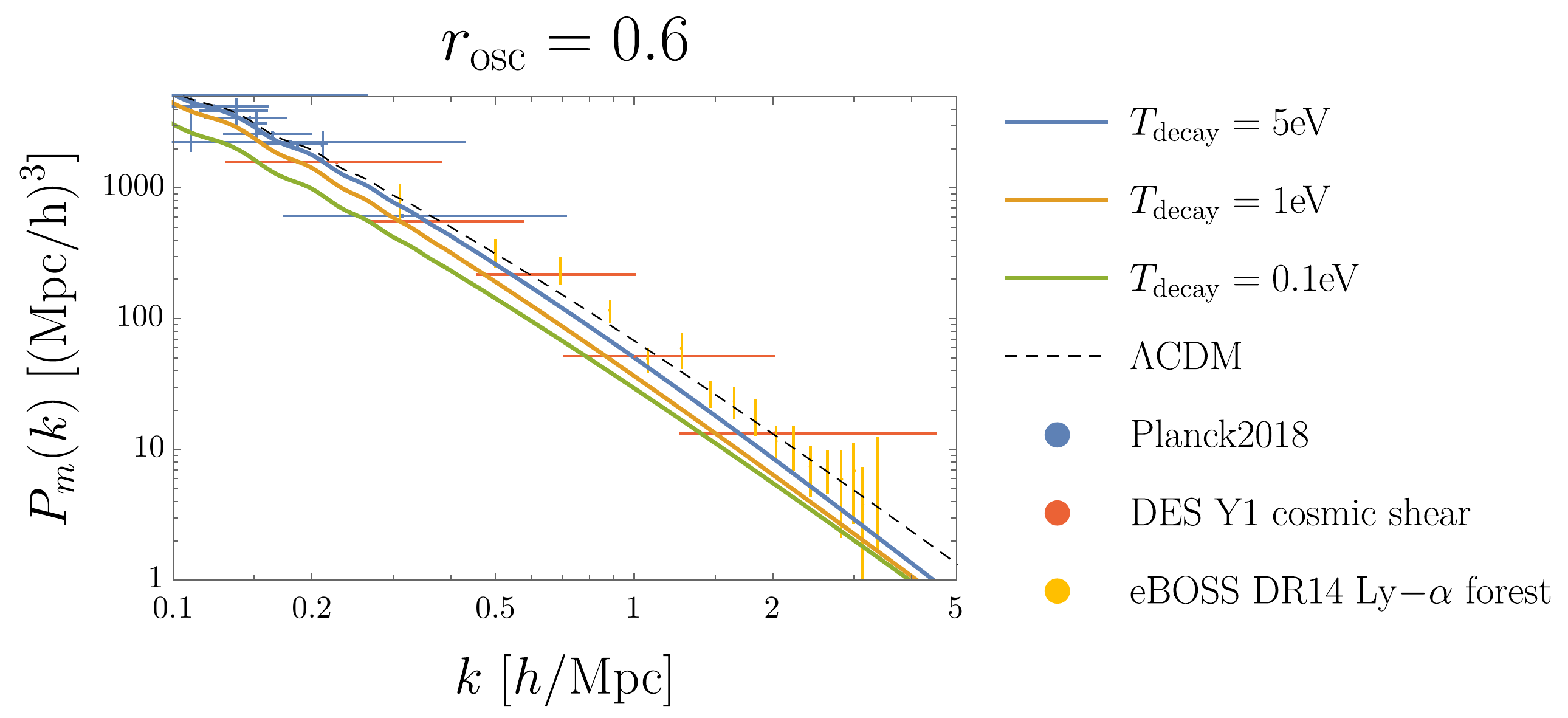}
        \caption{
        Same as Fig.~\ref{matter_power_spectrum1} but for the range $0.1 < k \, [h \mathrm{ /Mpc] }< 5$.
        }
        \label{matter_power_spectrum1_enlarge}
    \end{center}
\end{figure}

\begin{figure}[htbp]
    \begin{center}
        \includegraphics[clip,width=14cm]{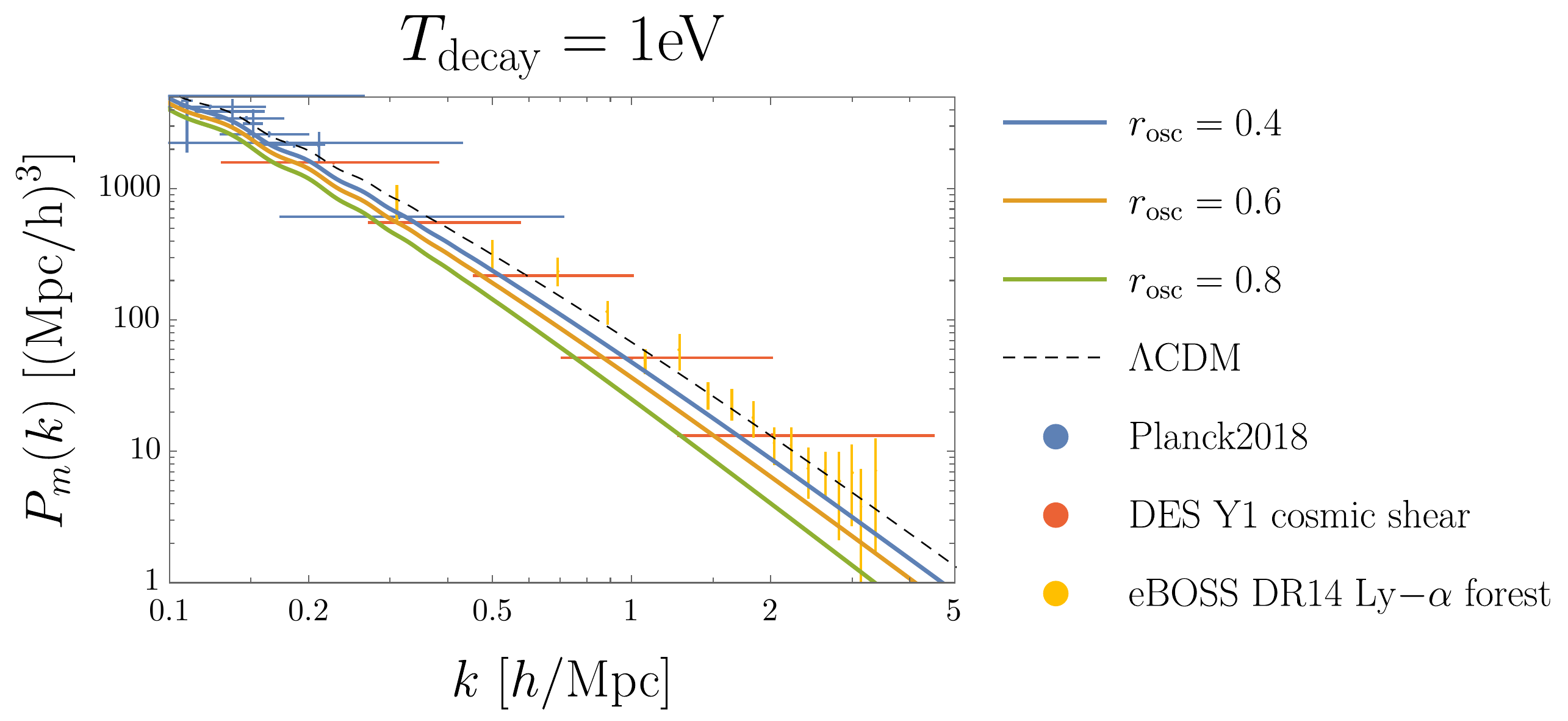}
        \caption{
        Matter power spectrum with $T_{ \mathrm{decay} } = 1$~eV. 
        The blue, orange and green solid lines represent the matter power spectra with $r_{ \mathrm{osc} } =$ 0.4, 0.6 and 0.8 respectively. The other explanations are the same as Figs.~\ref{matter_power_spectrum1},~\ref{matter_power_spectrum1_enlarge}. 
        }
        \label{matter_power_spectrum2}
    \end{center}
\end{figure}

Figs.~\ref{matter_power_spectrum1} and \ref{matter_power_spectrum1_enlarge} show the matter power spectra for $r_{ \mathrm{osc} } = 0.6$~\cite{Kawasaki:2020jnw} and $T_\text{decay}=5$\,eV, $1$\,eV, $0.1$\,eV.
From these figures, we can see that the matter power spectrum is inconsistent with the observed data (especially Lyman-$\alpha$ forest) for $T_\text{decay}\lesssim 1$~eV.
Fig.~\ref{matter_power_spectrum2} shows the matter power spectrum for $T_{ \mathrm{decay} } = 1$~eV and $r_{ \mathrm{osc} } =$0.4, 0.6, and 0.8.
From this figure, it can be seen that the larger the value of $r_{ \mathrm{osc} }$ is, the less it fits the observations.


We comment on the values of $(a,b,c)$~[Eq.~\eqref{N body parameter}], which is determined for the thermally produced warm dark matter model~~\cite{Kamada:2016vsc}.
These values might be changed for our case since $k'_d$ differently depends on the density ratio of the warm component, $k'_d\propto (\Omega_\mathrm{warm}/\Omega_\mathrm{DM})^{-5/6}$ for thermally produced warm dark matter and $k'_d\propto r_\mathrm{osc}^{-1/2}$ for our case.
We check the robustness of our results over the different values of Eq.~\eqref{N body parameter}.
We compute the matter power spectra for the respective powers of $k'_d \propto r_\mathrm{osc}^{-5/6}$ and $\propto r_\mathrm{osc}^{-1/2}$ using the same parameter of Eq.~\eqref{N body parameter} and the result is not changed, that is the oscillon decay temperature below $\mathcal{O}$(1)~eV is inconsistent with observation.

\begin{figure}[t]
    \begin{center}
        \includegraphics[clip,width=14cm]{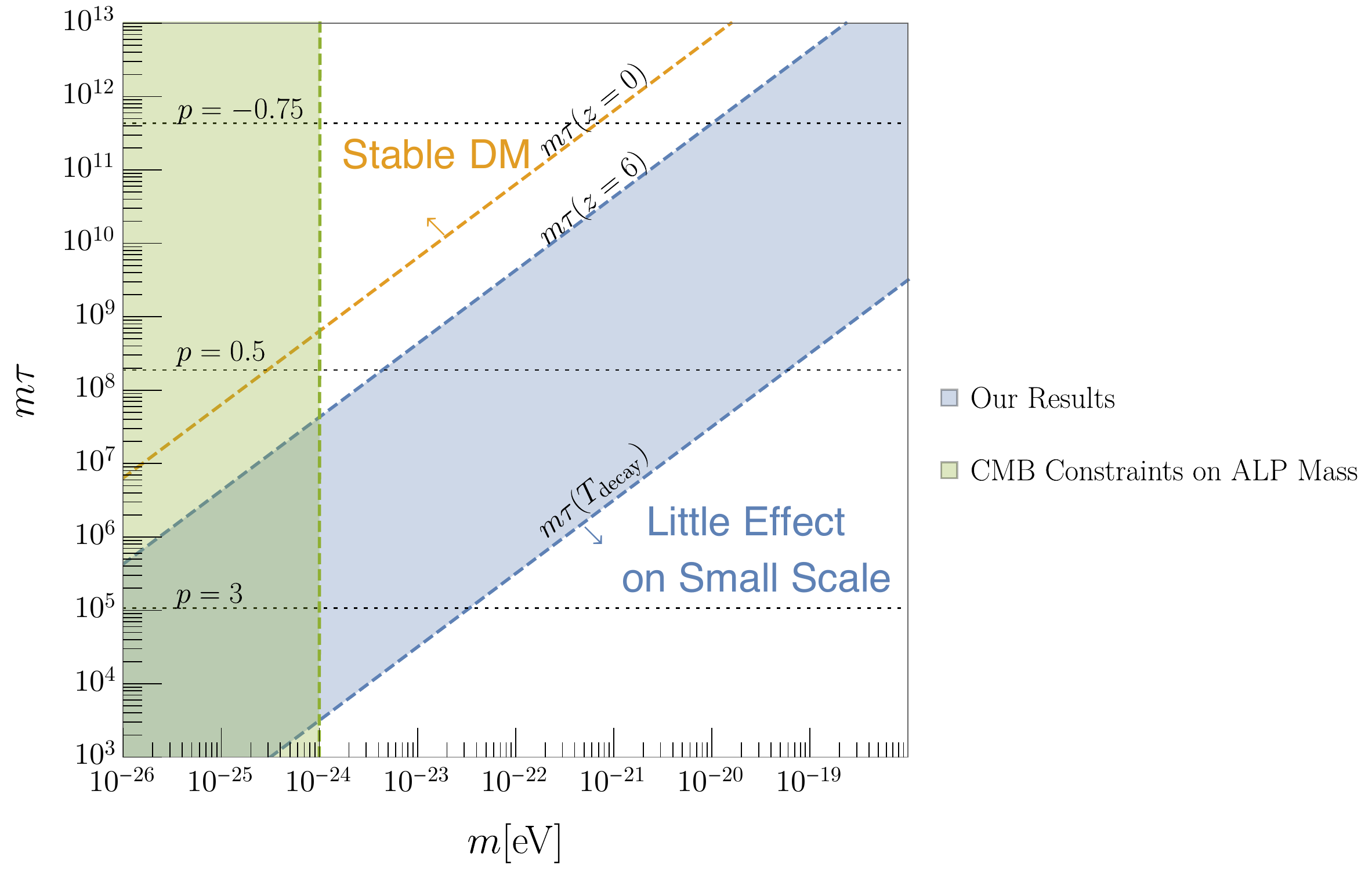}
        \caption{
        Constraints on the lifetime of oscillons $m \tau$ and ALP mass with $r_{ \mathrm{osc} } = 0.6$.
        The blue and orange dashed lines represent the lifetimes of oscillons that decay at $z=0$, $6$ and $T _{ \mathrm{decay} } = 1$~eV, respectively.
        The blue region is excluded because the free streaming of ALPs is too large to explain the small-scale structures as shown in Figs.~\ref{matter_power_spectrum1} and \ref{matter_power_spectrum2}.
        The green region is excluded because of the CMB observations~\cite{Hlozek:2014lca}, in which it is assumed that ALPs constitute 100\% of dark matter.
        The black dotted lines are the typical lifetime of oscillons with the potential~\eqref{ALPpot} at $p=-0.75, 0.5$ and $3$ respectively~\cite{Kawasaki:2020jnw}.
        In the upper left region labeled ``Stable Oscillon", oscillons have a long lifetime and remain in the current universe.
        In the lower right region labeled ``Little Effect on Small Scale", the lifetime of oscillons is too short and the FSL of the ALPs is too short to have any effect on the small scale structures.
        }
        \label{constraints}
    \end{center}
\end{figure}

Fig.~\ref{constraints} shows the observational constraints on the lifetime of oscillons and ALP mass from the matter power spectrum. 
The blue shaded region in Fig.~\ref{constraints} is excluded because the free-streaming length of the radiated ALPs is too large to form the small-scale structures.
The ALP with mass smaller than about $10^{-24}$~eV (green shaded region) is excluded from the CMB observations~\cite{Hlozek:2014lca}.
The black dotted lines show the numerically calculated lifetimes of the oscillon for the potential~\eqref{ALPpotential}~\cite{Kawasaki:2020jnw}. 
From Fig.~\ref{constraints}, the theoretical lifetimes corresponding to various values of $p$ in the potential~\eqref{ALPpotential} are limited.

\section{Conclusion}
\label{Conclusion}

In this paper, we studied the decay of oscillons produced from ALPs and estimated the FSL of ALPs emitted from the decay. 
Since the long FSL suppresses small-scale density perturbations, we can obtain constraints on the ALP mass and decay epoch of the oscillons from the small-scale observations. 
We have found that the matter power spectrum contradicts the observational data such as the Lyman-$\alpha$ forest if the ALP oscillons decay at the cosmic temperature below $1$~eV, from which  we obtained a limit on the mass of the ALPs.

In this paper, we adopt the pure natural type~\eqref{ALPpot} as the ALP potential, which leads to long lifetimes $\tau \gg 10^{3}m^{-1}$ and satisfies Eq.~\eqref{FSvsExpansion1}.
On the other hand, the cosine potential that is often used for simplicity predicts shorter lifetimes than $10^{3}m^{-1}$~\cite{Zhang:2020bec}. 
In this case, as suggested from Fig.~\ref{constraints}, the oscillon formation does not affect the small-scale structure of the universe.

The value of $\bar{k}$ is obtained for the potential~\eqref{ALPpot}.
In addition, we confirmed that the order of the value of $\bar{k}$ is almost the same ($\bar{k} \simeq 0.1 m$) for the $\phi ^6$ potential which is used in~\cite{Ibe:2019vyo,Mukaida:2016hwd}.
For simplicity, we assume that all oscillons have the same lifetime, but in reality, oscillons with various lifetimes are formed.
The effect of ALPs on the matter power spectrum in such a system will be investigated in the future.

Note that we changed the $r_{ \mathrm{osc} }$-dependence of $k'_{d}$ in Eq.~\eqref{damping_scale} from the original paper~\cite{Kamada:2016vsc} due to physical requirements, and this should also change the fitting parameters $(a, b, c)$ of Eq.~\eqref{N body parameter}.
Although the usage of the fitting parameters~\cite{Kamada:2016vsc} is justified as in Sec.~\ref{Matter Power Spectrum of Axion-like particles}, in order to determine these parameters accurately, N-body calculations are required.

We assume the only particles that affect the free-streaming are those emitted from the oscillon decays.
In fact, ALPs radiated before the decay of oscillon are less likely to affect the FSL because they lose momentum more rapidly due to the larger effect of cosmic expansion.
However the effect of evaporation to reduce $r_{ \mathrm{osc} }$ is not negligible and the effect of free-streaming particles emitted by evaporation should also be considered for a more detailed analysis.

\section*{Acknowledgments}

This work is supported by the Grant-in-Aid for Scientific Research Fund of the JSPS 20H05851(M.\,K.), 21K03567(M.\,K.),
20J20248 (K.\,M.), and
19J21974 (H.\,N.).
M.\,K. and K.\,M. are supported by World Premier International Research Center Initiative (WPI Initiative), MEXT, Japan (M.\,K. and K.\,M.). 
K.\,M. is supported by the Program of Excellence in Photon Science.
H.\,N. is supported by Advanced Leading Graduate Course for Photon Science.

\small
\bibliographystyle{JHEP}
\bibliography{bibtex}

\providecommand{\href}[2]{#2}\begingroup\raggedright\begin{thebibliography}{10}

\bibitem{Hinshaw_2013}
G.~Hinshaw, D.~Larson, E.~Komatsu, D.N.~Spergel, C.L.~Bennett, J.~Dunkley
  et~al., \emph{Nine-year wilkinson microwave anisotropy probe ( wmap )
  observations: Cosmological parameter results},
  \href{https://doi.org/10.1088/0067-0049/208/2/19}{\emph{The Astrophysical
  Journal Supplement Series} {\bfseries 208} (2013) 19}.

\bibitem{DES:2017myr}
{\scshape DES} collaboration, \emph{{Dark Energy Survey year 1 results:
  Cosmological constraints from galaxy clustering and weak lensing}},
  \href{https://doi.org/10.1103/PhysRevD.98.043526}{\emph{Phys. Rev. D}
  {\bfseries 98} (2018) 043526}
  [\href{https://arxiv.org/abs/1708.01530}{{\ttfamily 1708.01530}}].

\bibitem{Planck:2018vyg}
{\scshape Planck} collaboration, \emph{{Planck 2018 results. VI. Cosmological
  parameters}},
  \href{https://doi.org/10.1051/0004-6361/201833910}{\emph{Astron. Astrophys.}
  {\bfseries 641} (2020) A6}
  [\href{https://arxiv.org/abs/1807.06209}{{\ttfamily 1807.06209}}].

\bibitem{Moore:1999nt}
B.~Moore, S.~Ghigna, F.~Governato, G.~Lake, T.R.~Quinn, J.~Stadel et~al.,
  \emph{{Dark matter substructure within galactic halos}},
  \href{https://doi.org/10.1086/312287}{\emph{Astrophys. J. Lett.} {\bfseries
  524} (1999) L19} [\href{https://arxiv.org/abs/astro-ph/9907411}{{\ttfamily
  astro-ph/9907411}}].

\bibitem{de_Blok_2010}
W.J.G.~de~Blok, \emph{{The Core-Cusp Problem}},
  \href{https://doi.org/10.1155/2010/789293}{\emph{Advances in Astronomy}
  {\bfseries 2010} (2010) 1–14}.

\bibitem{Boylan_Kolchin_2012}
M.~Boylan-Kolchin, J.S.~Bullock and M.~Kaplinghat, \emph{{The Milky Way’s
  bright satellites as an apparent failure of $\Lambda$CDM}},
  \href{https://doi.org/10.1111/j.1365-2966.2012.20695.x}{\emph{Monthly Notices
  of the Royal Astronomical Society} {\bfseries 422} (2012) 1203–1218}.

\bibitem{Bullock:2017xww}
J.S.~Bullock and M.~Boylan-Kolchin, \emph{{Small-Scale Challenges to the
  $\Lambda$CDM Paradigm}},
  \href{https://doi.org/10.1146/annurev-astro-091916-055313}{\emph{Ann. Rev.
  Astron. Astrophys.} {\bfseries 55} (2017) 343}
  [\href{https://arxiv.org/abs/1707.04256}{{\ttfamily 1707.04256}}].

\bibitem{Hu:2000ke}
W.~Hu, R.~Barkana and A.~Gruzinov, \emph{{Cold and fuzzy dark matter}},
  \href{https://doi.org/10.1103/PhysRevLett.85.1158}{\emph{Phys. Rev. Lett.}
  {\bfseries 85} (2000) 1158}
  [\href{https://arxiv.org/abs/astro-ph/0003365}{{\ttfamily
  astro-ph/0003365}}].

\bibitem{Hui:2016ltb}
L.~Hui, J.P.~Ostriker, S.~Tremaine and E.~Witten, \emph{{Ultralight scalars as
  cosmological dark matter}},
  \href{https://doi.org/10.1103/PhysRevD.95.043541}{\emph{Phys. Rev. D}
  {\bfseries 95} (2017) 043541}
  [\href{https://arxiv.org/abs/1610.08297}{{\ttfamily 1610.08297}}].

\bibitem{Svrcek:2006yi}
P.~Svrcek and E.~Witten, \emph{{Axions In String Theory}},
  \href{https://doi.org/10.1088/1126-6708/2006/06/051}{\emph{JHEP} {\bfseries
  06} (2006) 051} [\href{https://arxiv.org/abs/hep-th/0605206}{{\ttfamily
  hep-th/0605206}}].

\bibitem{Dong:2010in}
X.~Dong, B.~Horn, E.~Silverstein and A.~Westphal, \emph{{Simple exercises to
  flatten your potential}},
  \href{https://doi.org/10.1103/PhysRevD.84.026011}{\emph{Phys. Rev. D}
  {\bfseries 84} (2011) 026011}
  [\href{https://arxiv.org/abs/1011.4521}{{\ttfamily 1011.4521}}].

\bibitem{Kallosh:2013hoa}
R.~Kallosh and A.~Linde, \emph{{Universality Class in Conformal Inflation}},
  \href{https://doi.org/10.1088/1475-7516/2013/07/002}{\emph{JCAP} {\bfseries
  07} (2013) 002} [\href{https://arxiv.org/abs/1306.5220}{{\ttfamily
  1306.5220}}].

\bibitem{Kallosh:2013yoa}
R.~Kallosh, A.~Linde and D.~Roest, \emph{{Superconformal Inflationary
  $\alpha$-Attractors}},
  \href{https://doi.org/10.1007/JHEP11(2013)198}{\emph{JHEP} {\bfseries 11}
  (2013) 198} [\href{https://arxiv.org/abs/1311.0472}{{\ttfamily 1311.0472}}].

\bibitem{Gleiser:1993pt}
M.~Gleiser, \emph{{Pseudostable bubbles}},
  \href{https://doi.org/10.1103/PhysRevD.49.2978}{\emph{Phys. Rev. D}
  {\bfseries 49} (1994) 2978}
  [\href{https://arxiv.org/abs/hep-ph/9308279}{{\ttfamily hep-ph/9308279}}].

\bibitem{Copeland:1995fq}
E.J.~Copeland, M.~Gleiser and H.R.~Muller, \emph{{Oscillons: Resonant
  configurations during bubble collapse}},
  \href{https://doi.org/10.1103/PhysRevD.52.1920}{\emph{Phys. Rev. D}
  {\bfseries 52} (1995) 1920}
  [\href{https://arxiv.org/abs/hep-ph/9503217}{{\ttfamily hep-ph/9503217}}].

\bibitem{Amin:2011hj}
M.A.~Amin, R.~Easther, H.~Finkel, R.~Flauger and M.P.~Hertzberg,
  \emph{{Oscillons After Inflation}},
  \href{https://doi.org/10.1103/PhysRevLett.108.241302}{\emph{Phys. Rev. Lett.}
  {\bfseries 108} (2012) 241302}
  [\href{https://arxiv.org/abs/1106.3335}{{\ttfamily 1106.3335}}].

\bibitem{Amin:2010dc}
M.A.~Amin, R.~Easther and H.~Finkel, \emph{{Inflaton Fragmentation and Oscillon
  Formation in Three Dimensions}},
  \href{https://doi.org/10.1088/1475-7516/2010/12/001}{\emph{JCAP} {\bfseries
  12} (2010) 001} [\href{https://arxiv.org/abs/1009.2505}{{\ttfamily
  1009.2505}}].

\bibitem{Amin:2019ums}
M.A.~Amin and P.~Mocz, \emph{{Formation, gravitational clustering, and
  interactions of nonrelativistic solitons in an expanding universe}},
  \href{https://doi.org/10.1103/PhysRevD.100.063507}{\emph{Phys. Rev. D}
  {\bfseries 100} (2019) 063507}
  [\href{https://arxiv.org/abs/1902.07261}{{\ttfamily 1902.07261}}].

\bibitem{Lozanov:2017hjm}
K.D.~Lozanov and M.A.~Amin, \emph{{Self-resonance after inflation: oscillons,
  transients and radiation domination}},
  \href{https://doi.org/10.1103/PhysRevD.97.023533}{\emph{Phys. Rev. D}
  {\bfseries 97} (2018) 023533}
  [\href{https://arxiv.org/abs/1710.06851}{{\ttfamily 1710.06851}}].

\bibitem{Kitajima:2018zco}
N.~Kitajima, J.~Soda and Y.~Urakawa, \emph{{Gravitational wave forest from
  string axiverse}},
  \href{https://doi.org/10.1088/1475-7516/2018/10/008}{\emph{JCAP} {\bfseries
  10} (2018) 008} [\href{https://arxiv.org/abs/1807.07037}{{\ttfamily
  1807.07037}}].

\bibitem{Hong:2017ooe}
J.-P.~Hong, M.~Kawasaki and M.~Yamazaki, \emph{{Oscillons from Pure Natural
  Inflation}}, \href{https://doi.org/10.1103/PhysRevD.98.043531}{\emph{Phys.
  Rev. D} {\bfseries 98} (2018) 043531}
  [\href{https://arxiv.org/abs/1711.10496}{{\ttfamily 1711.10496}}].

\bibitem{Ibe:2019vyo}
M.~Ibe, M.~Kawasaki, W.~Nakano and E.~Sonomoto, \emph{{Decay of I-ball/Oscillon
  in Classical Field Theory}},
  \href{https://doi.org/10.1007/JHEP04(2019)030}{\emph{JHEP} {\bfseries 04}
  (2019) 030} [\href{https://arxiv.org/abs/1901.06130}{{\ttfamily
  1901.06130}}].

\bibitem{Zhang:2020bec}
H.-Y.~Zhang, M.A.~Amin, E.J.~Copeland, P.M.~Saffin and K.D.~Lozanov,
  \emph{{Classical Decay Rates of Oscillons}},
  \href{https://doi.org/10.1088/1475-7516/2020/07/055}{\emph{JCAP} {\bfseries
  07} (2020) 055} [\href{https://arxiv.org/abs/2004.01202}{{\ttfamily
  2004.01202}}].

\bibitem{Kawasaki:2020jnw}
M.~Kawasaki, W.~Nakano, H.~Nakatsuka and E.~Sonomoto, \emph{{Oscillons of
  Axion-Like Particle: Mass distribution and power spectrum}},
  \href{https://doi.org/10.1088/1475-7516/2021/01/061}{\emph{JCAP} {\bfseries
  01} (2021) 061} [\href{https://arxiv.org/abs/2010.09311}{{\ttfamily
  2010.09311}}].

\bibitem{Kasuya:2002zs}
S.~Kasuya, M.~Kawasaki and F.~Takahashi, \emph{{I-balls}},
  \href{https://doi.org/10.1016/S0370-2693(03)00344-7}{\emph{Phys. Lett. B}
  {\bfseries 559} (2003) 99}
  [\href{https://arxiv.org/abs/hep-ph/0209358}{{\ttfamily hep-ph/0209358}}].

\bibitem{Kawasaki:2015vga}
M.~Kawasaki, F.~Takahashi and N.~Takeda, \emph{{Adiabatic Invariance of
  Oscillons/I-balls}},
  \href{https://doi.org/10.1103/PhysRevD.92.105024}{\emph{Phys. Rev. D}
  {\bfseries 92} (2015) 105024}
  [\href{https://arxiv.org/abs/1508.01028}{{\ttfamily 1508.01028}}].

\bibitem{Silverstein:2008sg}
E.~Silverstein and A.~Westphal, \emph{{Monodromy in the CMB: Gravity Waves and
  String Inflation}},
  \href{https://doi.org/10.1103/PhysRevD.78.106003}{\emph{Phys. Rev. D}
  {\bfseries 78} (2008) 106003}
  [\href{https://arxiv.org/abs/0803.3085}{{\ttfamily 0803.3085}}].

\bibitem{McAllister:2008hb}
L.~McAllister, E.~Silverstein and A.~Westphal, \emph{{Gravity Waves and Linear
  Inflation from Axion Monodromy}},
  \href{https://doi.org/10.1103/PhysRevD.82.046003}{\emph{Phys. Rev. D}
  {\bfseries 82} (2010) 046003}
  [\href{https://arxiv.org/abs/0808.0706}{{\ttfamily 0808.0706}}].

\bibitem{Nomura:2017ehb}
Y.~Nomura, T.~Watari and M.~Yamazaki, \emph{{Pure Natural Inflation}},
  \href{https://doi.org/10.1016/j.physletb.2017.11.052}{\emph{Phys. Lett. B}
  {\bfseries 776} (2018) 227}
  [\href{https://arxiv.org/abs/1706.08522}{{\ttfamily 1706.08522}}].

\bibitem{Kawasaki:2019czd}
M.~Kawasaki, W.~Nakano and E.~Sonomoto, \emph{{Oscillon of Ultra-Light
  Axion-like Particle}},
  \href{https://doi.org/10.1088/1475-7516/2020/01/047}{\emph{JCAP} {\bfseries
  01} (2020) 047} [\href{https://arxiv.org/abs/1909.10805}{{\ttfamily
  1909.10805}}].

\bibitem{Mukaida:2016hwd}
K.~Mukaida, M.~Takimoto and M.~Yamada, \emph{{On Longevity of
  I-ball/Oscillon}}, \href{https://doi.org/10.1007/JHEP03(2017)122}{\emph{JHEP}
  {\bfseries 03} (2017) 122}
  [\href{https://arxiv.org/abs/1612.07750}{{\ttfamily 1612.07750}}].

\bibitem{Eby:2018ufi}
J.~Eby, K.~Mukaida, M.~Takimoto, L.C.R.~Wijewardhana and M.~Yamada,
  \emph{{Classical nonrelativistic effective field theory and the role of
  gravitational interactions}},
  \href{https://doi.org/10.1103/PhysRevD.99.123503}{\emph{Phys. Rev. D}
  {\bfseries 99} (2019) 123503}
  [\href{https://arxiv.org/abs/1807.09795}{{\ttfamily 1807.09795}}].

\bibitem{Salmi:2012ta}
P.~Salmi and M.~Hindmarsh, \emph{{Radiation and Relaxation of Oscillons}},
  \href{https://doi.org/10.1103/PhysRevD.85.085033}{\emph{Phys. Rev. D}
  {\bfseries 85} (2012) 085033}
  [\href{https://arxiv.org/abs/1201.1934}{{\ttfamily 1201.1934}}].

\bibitem{Kamada:2016vsc}
A.~Kamada, K.T.~Inoue and T.~Takahashi, \emph{{Constraints on mixed dark matter
  from anomalous strong lens systems}},
  \href{https://doi.org/10.1103/PhysRevD.94.023522}{\emph{Phys. Rev. D}
  {\bfseries 94} (2016) 023522}
  [\href{https://arxiv.org/abs/1604.01489}{{\ttfamily 1604.01489}}].

\bibitem{Inoue:2014jka}
K.T.~Inoue, R.~Takahashi, T.~Takahashi and T.~Ishiyama, \emph{{Constraints on
  warm dark matter from weak lensing in anomalous quadruple lenses}},
  \href{https://doi.org/10.1093/mnras/stv194}{\emph{Mon. Not. Roy. Astron.
  Soc.} {\bfseries 448} (2015) 2704}
  [\href{https://arxiv.org/abs/1409.1326}{{\ttfamily 1409.1326}}].

\bibitem{Lewis:1999bs}
A.~Lewis, A.~Challinor and A.~Lasenby, \emph{{Efficient computation of CMB
  anisotropies in closed FRW models}},
  \href{https://doi.org/10.1086/309179}{\emph{Astrophys. J.} {\bfseries 538}
  (2000) 473} [\href{https://arxiv.org/abs/astro-ph/9911177}{{\ttfamily
  astro-ph/9911177}}].

\bibitem{DES:2017qwj}
{\scshape DES} collaboration, \emph{{Dark Energy Survey Year 1 results:
  Cosmological constraints from cosmic shear}},
  \href{https://doi.org/10.1103/PhysRevD.98.043528}{\emph{Phys. Rev. D}
  {\bfseries 98} (2018) 043528}
  [\href{https://arxiv.org/abs/1708.01538}{{\ttfamily 1708.01538}}].

\bibitem{Chabanier:2019eai}
S.~Chabanier, M.~Millea and N.~Palanque-Delabrouille, \emph{{Matter power
  spectrum: from Ly$\alpha$ forest to CMB scales}},
  \href{https://doi.org/10.1093/mnras/stz2310}{\emph{Mon. Not. Roy. Astron.
  Soc.} {\bfseries 489} (2019) 2247}
  [\href{https://arxiv.org/abs/1905.08103}{{\ttfamily 1905.08103}}].

\bibitem{2018ApJS..235...42A}
B.~{Abolfathi}, D.S.~{Aguado}, G.~{Aguilar}, C.~{Allende Prieto}, A.~{Almeida},
  T.T.~{Ananna} et~al., \emph{{The Fourteenth Data Release of the Sloan Digital
  Sky Survey: First Spectroscopic Data from the Extended Baryon Oscillation
  Spectroscopic Survey and from the Second Phase of the Apache Point
  Observatory Galactic Evolution Experiment}},
  \href{https://doi.org/10.3847/1538-4365/aa9e8a}{\emph{Astrophys. J. Suppl.}
  {\bfseries 235} (2018) 42}
  [\href{https://arxiv.org/abs/1707.09322}{{\ttfamily 1707.09322}}].

\bibitem{Hlozek:2014lca}
R.~Hlozek, D.~Grin, D.J.E.~Marsh and P.G.~Ferreira, \emph{{A search for
  ultralight axions using precision cosmological data}},
  \href{https://doi.org/10.1103/PhysRevD.91.103512}{\emph{Phys. Rev. D}
  {\bfseries 91} (2015) 103512}
  [\href{https://arxiv.org/abs/1410.2896}{{\ttfamily 1410.2896}}].

\end{thebibliography}\endgroup

\end{document}